\documentclass[%
 reprint,
 amsmath,amssymb,
 aps,
prl
]{revtex4-2}

\usepackage{subcaption}
\usepackage{xr}
\usepackage{accents}
\usepackage{comment}
\usepackage{booktabs}
\usepackage{graphicx}
\usepackage{dcolumn}
\usepackage{bm}
\usepackage{xcolor}
\usepackage{multirow}
\usepackage{xr}

\begin{document}


\title{Data-driven modeling of multiscale phenomena with applications to fluid turbulence}

\author{Brandon Choi$^1$}
\author{Matteo Ugliotti$^1$}
\author{Mateo Reynoso$^1$}
\author{Daniel R. Gurevich$^2$}
\author{Roman O. Grigoriev$^1$}
 \email{roman.grigoriev@physics.gatech.edu}
\affiliation{%
$^1$School of Physics, Georgia Institute of Technology, Atlanta, GA 30332, USA
}%
\affiliation{%
$^2$Department of Mathematics, University of California, Los Angeles, CA 90095, USA
}%

\date{\today}

\begin{abstract}
This paper introduces a novel data-driven framework for constructing accurate and general equivariant models of multiscale phenomena which does not rely on specific assumptions about the underlying physics. This framework is illustrated using incompressible fluid turbulence as an example that is representative, practically important, reasonably simple, and exceedingly well-studied. We use direct numerical simulations of freely decaying turbulence in two spatial dimensions to infer an effective field theory comprising explicit, interpretable evolution equations for both the large (resolved) and small (modeled) scales. The resulting closed system of equations is capable of accurately describing the effect of small scales, including backscatter---the flow of energy from small to large scales, which is particularly pronounced in two dimensions---which is an outstanding challenge that, to our knowledge, no existing alternative successfully tackles.
\end{abstract}

\maketitle

\section{Introduction}
Construction of mathematical models of physical systems---at length, time and, energy scales of interest---has long stood at the foundation of physics. This cornerstone paved the way to improving our understanding of the physical world and has made numerous groundbreaking predictions. Often, a {\it fundamental theory} is available which provides an exact description of the microscopic details but is practically intractable or fails to generate the needed insight at the scales of interest. In such cases, it is desirable to construct an {\it effective field theory} (EFT) that only describes the relevant degrees of freedom at the scales of interest without trying to model microscopic details. 
Effective theories are commonly used in particle, nuclear \cite{hammer2020}, and condensed matter physics \cite{fradkin2013}, but examples such as the Chapman-Enskog theory \cite{chapman1990} are found in classical physics as well. 

A top-down effective theory can be derived formally from the corresponding fundamental theory only in very specific limits, e.g., the Chapman-Enskog theory requires the gas to be near local thermodynamic equilibrium and the Knudsen number (characterizing the mean free path) to be small. However, when formal derivations fail, one can ask whether a sufficiently general approach exists that, at least in some classes of problems, allows an effective theory to be inferred in a systematic way from the corresponding fundamental theory. A particular class of problems where this capability would be most welcome involves multiscale phenomena, such as turbulence in fluids and plasmas, which can feature structures with length and time scales varying by tens of orders of magnitudes. 

Consider, for instance, Newtonian fluids which are well-described by the continuity and Navier-Stokes (momentum) equations 
\begin{subequations}\label{eq:fm}
\begin{align}
    \nabla\cdot{\bf u} &= 0, \label{eq:ic}\\
    \partial_t{\bf u} + {\bf u}\cdot\nabla {\bf u}&= -\nabla p+\nu \nabla^2 {\bf u}{\color{black}+{\bf f}}, \label{eq:NSE}
\end{align}
\end{subequations}
where $\nu$ is the kinematic viscosity, down to the scale of microns.
The momentum equation features a nonlinearity which severely complicates analysis and gives rise to the multiscale nature of turbulent flows  \cite{richardson1922}.
This multiscale nature of turbulence makes solution of the fundamental transport equations via
direct numerical simulations (DNS) infeasible in most practical applications. In particular, the ratio between the largest and smallest length scales in three-dimensional turbulence scales as $M\propto \mathrm{Re}^{3/4}$, where Re is the Reynolds number, while fully resolved DNS requires solving equations for $M^3$ degrees of freedom at each time step. For instance, atmospheric jet flows and oceanic currents are described by Re ranging between $10^{10}$ and $10^{12}$, which corresponds to $10^{22}\lesssim M^3\lesssim 10^{27}$. 

To circumvent this limitation, various models of fluid turbulence have been introduced which involve either temporal or spatial coarse-graining as in the Reynolds-averaged Navier-Stokes (RANS) \cite{alfonsi2009} or large eddy simulation (LES) \cite{meneveau2000,moser2021}, respectively. While the coarse-graining procedure itself is formal and general, the equations describing the averaged fields and their fluctuations cannot be closed without making specific assumptions, yielding a {\it phenomenological description}. In the case of both RANS and LES, an additional {\it closure} term appears in the momentum equation. It represents the effect of the small (modeled) scales on the large (resolved) ones and its functional form is assumed based on various empirical arguments rather than derived. 

For instance, in LES, the velocity and pressure fields are decomposed into the large- and small-scale components, $\phi = \bar \phi + \phi'$, using the filtering operator
\begin{equation} \label{filter}
    \overline{\phi}({\textbf{r}}) \equiv \int d\textbf{r}'\: G_\Delta(\textbf{r}-\textbf{r}')\phi(r'),
\end{equation}
where $\Delta$ is the cutoff length scale that defines the resolution of the resulting coarse-grained description. Applying this operator to the fundamental model \eqref{eq:fm}, one obtains a pair of equations for the large-scale components 
\begin{subequations}\label{eq:les}
\begin{align}
    \nabla_i \bar u_i &= 0, \\
    \partial_t \bar u_i + \bar u_j \nabla_j \bar u_i &= -\nabla_i \bar p+\nu \nabla^2 \bar u_i{\color{black}+\bar{\bf f}} - \nabla_j \tau_{ij}, \label{eq:FNSE}
\end{align}
\end{subequations}
where we use Einstein implicit summation notation.
The closure term $-\nabla_j \tau_{ij}$ is expressed via the subgrid-scale (SGS) stress tensor
$\tau_{ij} = \overline{u_i u_j} - \bar u_i \bar u_j$, which generally cannot be expressed in terms of the resolved fields (here, $\bar{\bf u}$ and $\bar{p}$) and thus requires modeling. 

A variety of phenomenological models, each with its strengths and weaknesses, have been proposed. The most widely used models are those based on the eddy viscosity hypothesis, 
such as the Smagorinsky model \cite{smagorinsky_general_1963} and its dynamic version \cite{germano_dynamic_1991,lilly_proposed_1992}.
Another popular type are the similarity models \cite{bardina_improved_1980,liu1994}. 
Models of both types, as well as hybrids thereof (e.g., the dynamic mixed model \cite{bardina_improved_1980}) perform well on some benchmarks \cite{vreman_large-eddy_1997}. However, they are all based on physical assumptions---such as statistical homogeneity, isotropy, and scale invariance---and struggle when these assumptions break, e.g., in the presence of coherent structures. 
Recent development of data-driven approaches has led to a flurry of efforts to construct closures as functions of the resolved fields by leveraging high-fidelity DNS data. However, only a small fraction of those studies were aimed at obtaining an explicit functional form of the relation, as opposed to black-box relations encoded by neural networks (see, e.g., \cite{kochkov2021}). Examples include approaches based on sparse regression (e.g., random forest regression \cite{ling2016}, sequential thresholding ridge regression \cite{schmelzer2018}, relevance vector machine \cite{jakhar2024}), symbolic regression (e.g., genetic expression programming \cite{weatheritt2017,reissmann2021}) or a combination of symbolic and sparse regression \cite{ross2023}. However, none of the data-driven models outperform the leading phenomenological models.

\begin{figure*}[htbp!]
    \centering
    \begin{subfigure}{0.28\textwidth}
    \includegraphics[width=\textwidth]{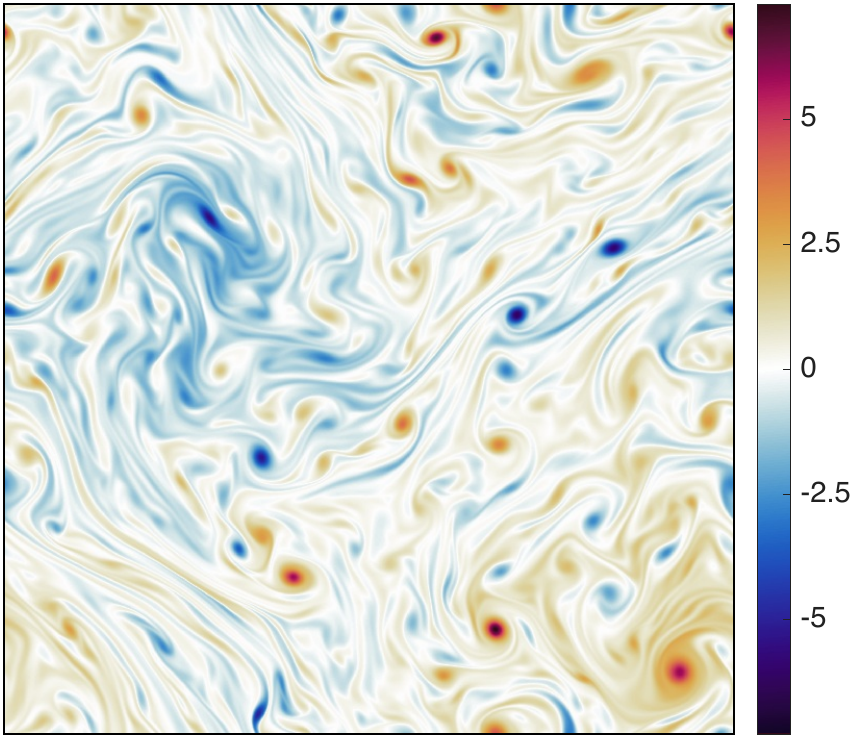}
    \caption*{(F1)}
    \end{subfigure}
    \hspace{1mm}
    \begin{subfigure}{0.275\textwidth}
    \includegraphics[width=\textwidth]{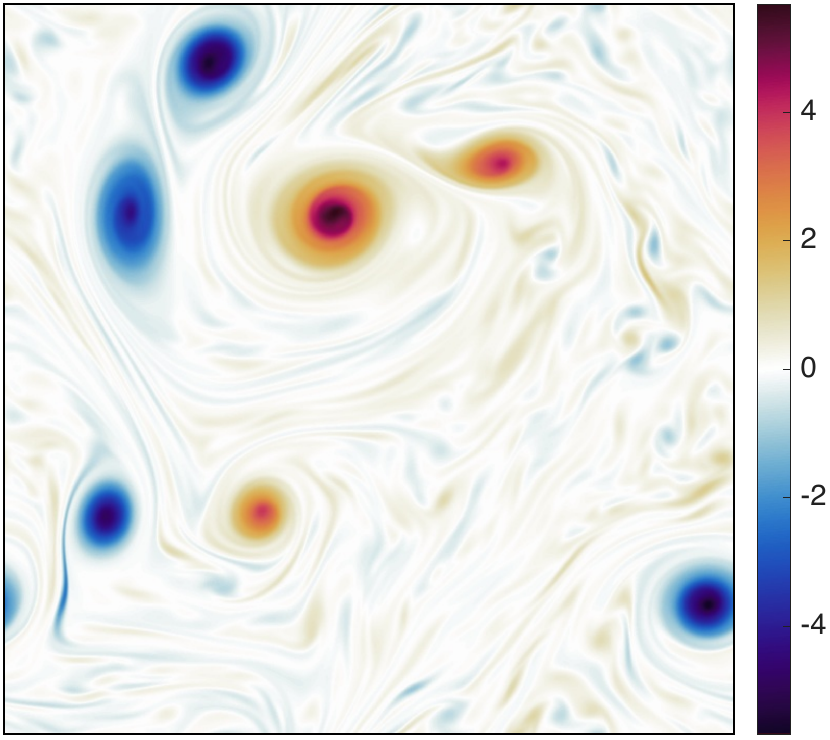}
    \caption*{(F2)}
    \end{subfigure}
    \hspace{1mm}
    \begin{subfigure}{0.28\textwidth}
    \includegraphics[width=\textwidth]{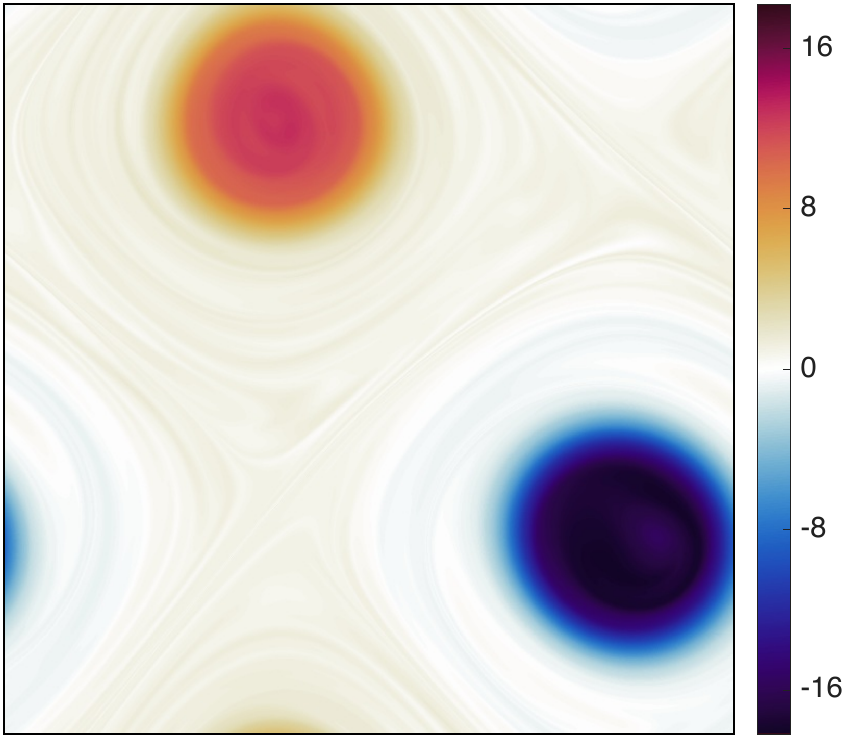}
    \caption*{(F3)}
    \end{subfigure}
    \hspace{1mm}
    \caption{The initial flow fields used to generate the training data. Shown is the initial vorticity field $\omega$ .}
    \label{fig:initial_conditions}
\end{figure*}

Coherent structures can be very prominent in some types of turbulent flows, e.g, those that are effectively two-dimensional. Such flows often feature strong local fluxes of energy 
from small to large scales. For a variety of reasons, no existing SGS models---phenomenological or data-driven---are capable of correctly describing backscatter, which plays a key role in a variety of physical phenomena. One prominent example is transport of angular momentum in accretion disks, where magneto-rotational instability that takes place at small scales substantially affects the large scales. Although efforts to develop an LES-like description of scale interaction in the context of magnetohydrodynamics have been made \cite{pessah2006}, proper treatment of backscatter in this and other contexts remains an open problem.

This paper introduces a general data-driven framework for inferring an EFT of multiscale phenomena from the underlying fundamental description. This framework requires no phenomenological assumptions and ensures that the resulting theory is equivariant and interpretable. To illustrate this framework, we apply it to fluid turbulence and infer a subgrid-scale model that accurately describes backscatter. This is due, in no small part, to the model's explicit characterization of the subgrid scales.

\section{Methodology}
We retain the coarse-graining approach used in LES to separate the degrees of freedom with length scales larger and smaller than a cutoff scale $\Delta$; while $\Delta$ is an adjustable parameter of the model, it must be at a scale substantially smaller than the scale of coherent structures, as SGS models are not designed to capture these. The choice of the kernel $G$ in equation \eqref{filter} is partly arbitrary but must satisfy a number of constraints such as locality in physical and Fourier space \cite{pope_turbulent_2000,sagaut_large_2006}. We use a normalized Gaussian kernel (with second moment $\sigma^2=\Delta^2/12$) exclusively in this paper as it maximizes locality in both spaces. To make the description finite-dimensional (e.g., to enable numerical solution), we additionally apply a sharp spectral filter, which cuts off scales past $\Delta$, making the combined filter non-invertible.
The resulting splitting formally defines the governing equations for both large scales (here the system \eqref{eq:les}) and small scales. The variables describing small scales and their associated governing equations are discarded in LES---which leads to the problems discussed previously---but are occasionally retained in RANS. In the latter case, the effect of small scales is represented by an SGS stress tensor whose formal evolution equation \cite{chou1945} includes variables other than $\bar{u}_i$, $\bar{p}$ and $\tau_{ij}$, generated by the nonlinearity, so that the resulting system of equations is not closed. 

Construction of a {\it closed} system of equations requires identifying (1) proper variables describing small scales, (2) approximate governing equations for these variables, and (3) an approximate constitutive relation for the SGS tensor.
While it is natural to expect the small scales to be described by additional tensor field(s) with their own evolution equation(s), the choice of these fields is not obvious {\it {\em a priori}}. 
The SGS tensor itself is generally a poor choice, as it also contains substantial contributions from the large scales. This can be seen easily by considering the exact Galilean-invariant decomposition \cite{clark_evaluation_1979,germano_proposal_1986}
\begin{subequations} \label{eq:decomp}
\begin{eqnarray}\label{eq:LCR}
   \tau_{ij} & = & L_{ij} + C_{ij} + R_{ij}
   \\[1pt]
   L_{ij} & = &  \overline{\overline{u}_i \overline{u}_j} - \overline{\overline{u}}_i  \overline{\overline{u}}_j 
   \\
  C_{ij} & = & \overline{\overline{u}_i u'_j + u'_i \overline{u}_j} -  \overline{\overline{u}}_i \overline{u'}_j -\overline{u'}_i \overline{\overline{u}}_j
   \\
  R_{ij} & = & \overline{ u'_i u'_j } - \overline{u'}_i \overline{u'}_j
\end{eqnarray}
\end{subequations}
of the SGS tensor into the Leonard ($L$), Reynolds ($R$), and cross ($C$) stress tensors describing, respectively, the interaction of large scales, interaction of small scales, and interscale interaction. While both ${\bf u}$ and ${\bf u}'$ (and thereby $L$, $C$ and $R$) can be reconstructed from $\bar{\bf u}$ exactly for the Gaussian filter,
the sharp spectral filter is not invertible and so some of the tensors may not be accurately reconstructed in terms of the resolved fields. 

On a sufficiently fine grid and for kernels with finite moments, any filtered field can be approximated using the \textit{moment expansion} \cite{sagaut_large_2006} as a series in $\Delta/\ell_c$, where $\ell_c$ is a characteristic length scale such as a vortex size. 
In particular, one finds $L=O((\Delta/\ell_c)^2)$, $C=O((\Delta/\ell_c)^4)$ and $R=O((\Delta/\ell_c)^6)$, which means that $L$ provides the dominant contribution to $\tau$ while $C$ and $R$ yield progressively smaller corrections. 
The leading order term of the moment expansion representation of $\tau$ (and $L$) yields the nonlinear gradient model \cite{leonard1999} that was recently rediscovered using a data-driven approach \cite{jakhar2024}. While this model yields a reasonably accurate approximation of the SGS tensor, it incorrectly predicts that the energy flux $\Pi \equiv -\tau_{ij}\,\bar S_{ij}$ (where $S_{ij}$ denotes the symmetric part of $\nabla_i u_j$) vanishes identically. Therefore, correct description of the fluxes require a far more accurate approximation of the SGS tensor.

We consider two-dimensional turbulence, which features pronounced coherent structures (eddies) that break statistical isotropy/homogeneity and is characterized by strong backscatter \cite{boffetta_two-dimensional_2011}, making it an ideal testing ground. We use a pseudospectral solver jax-cfd \cite{jaxcfd} to generate fully resolved DNS data on a square domain of size $\ell=2\pi$ with periodic boundary conditions.  
Figure \ref{fig:initial_conditions} shows snapshots of such flows representing the inverse cascade (F1), freely decaying turbulence (F2) and the direct cascade (F3) used as initial conditions. The procedure for generating these flows is described in the supplemental material \cite{supplemental_material}.
\textcolor{black}{It should be noted that, while the flow F1 does exhibit the condensate formation, the energy flux is not constant due to the absence of hypoviscosity. As a result, the inverse cascede is characterized by $k^{-3}$ rather than the classical $k^{-5/3}$ scaling \cite{zhu2024}. Similarly, the direct cascades in flows F1 and F3 feature a non-constant enstrophy flux and are characterized by power laws that are steeper than the classical prediction $k^{-3}$ \cite{reynoso2024}.}

Training data are generated as follows: solutions are computed without forcing (${\bf f}=0$) at viscosities $\nu=10^{-4}$, $10^{-5}$, and $10^{-6}$, on a computational grid with resolution $2048^2$, $4096^2$, and $4096^2$, respectively. {\color{black}Starting with each of the three initial conditions shown in figure \ref{fig:initial_conditions}, the flow is evolved over roughly one eddy turnover time. In total, this yields 9 numerical solutions.}
The corresponding Reynolds numbers $Re=u_{\rm RMS}\ell/\nu$ are reported in table \ref{tab:sample}. The initial conditions and the software for generating the data are available online \cite{sgs_github}. 

Next, all numerical solutions are coarse-grained by applying the filtering operator over a range of cutoff scales $\Delta$. From these coarse-grained datasets, the functional form of the SGS model is inferred using the Python implementation of SPIDER \cite{gurevich_phd,pyspider}. SPIDER combines group representation theory (to construct term libraries which define the search spaces), weak formulation of differential equations (to reduce symbolic relations involving derivatives to a system of linear equations for a set of coefficients), and sparse regression (to infer equivariant functional relations corresponding to different irreducible representations of the relevant symmetry group, in this case, SO(2)) \cite{golden_physically_2023,gurevich2024,wareing2025}. Finally, the relationship between the parameters of the EFT and the fundamental description (e.g., $\nu$) as well as the filter scale $\Delta$ are determined using a combination of dimensional analysis and scaling arguments.

\begin{figure*}[htbp!]
    \centering
    \begin{subfigure}{0.24\textwidth}
        \includegraphics[width=\textwidth]{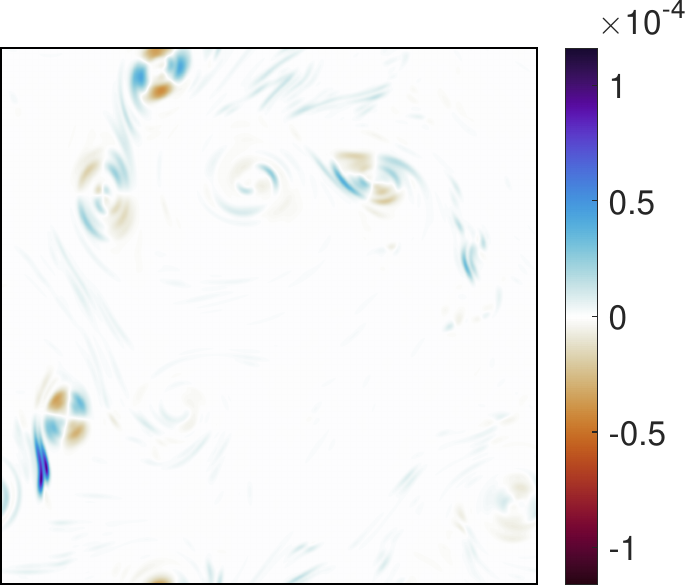}
    \caption*{(a)}
    \end{subfigure}
    \hspace{0mm}
    \begin{subfigure}{0.24\textwidth}
        \includegraphics[width=\textwidth]{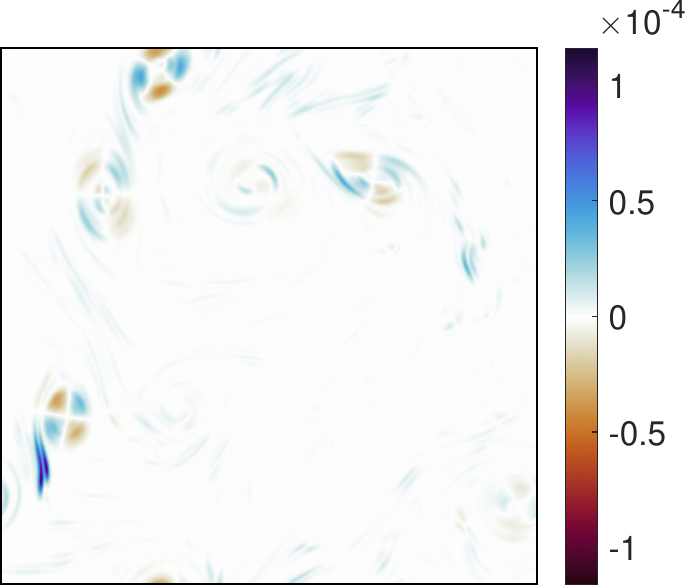}
    \caption*{(b)}
    \end{subfigure}
    \hspace{0mm}
    \begin{subfigure}{0.24\textwidth}
        \includegraphics[width=\textwidth]{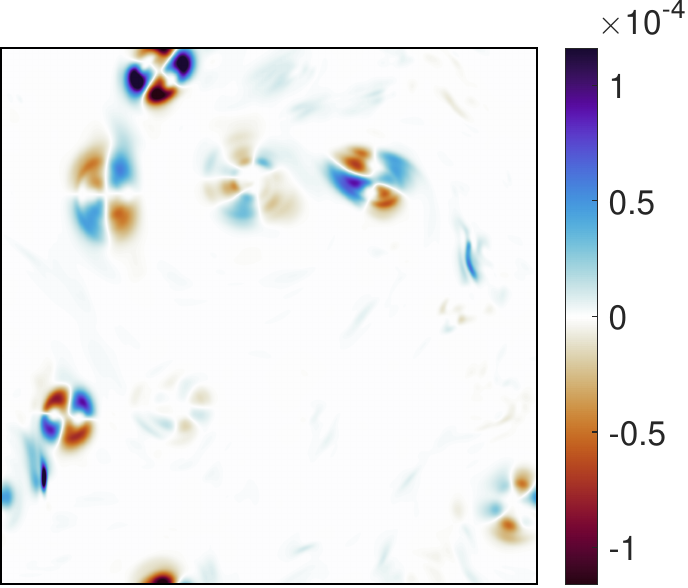}
    \caption*{(c)}
    \end{subfigure}
    \hspace{0mm}
    \begin{subfigure}{0.24\textwidth}
        \includegraphics[width=\textwidth]{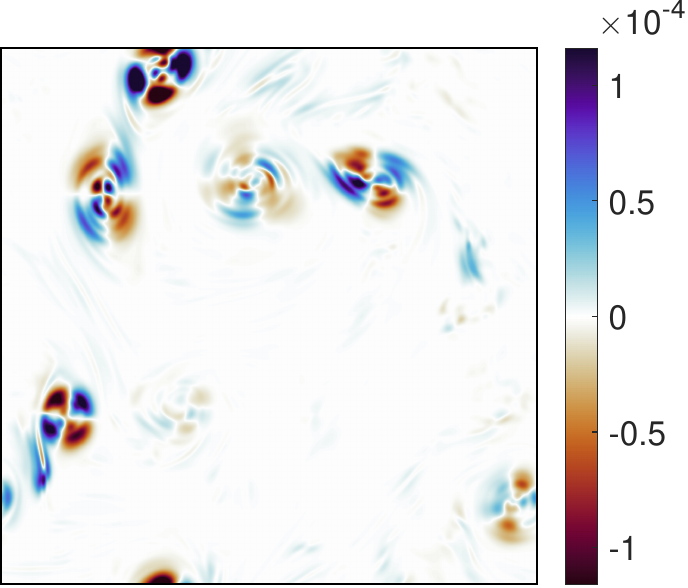}
    \caption*{(d)}
    \end{subfigure}
    \caption{The energy flux $\Pi$ of the flow F2 shown in Figure \ref{fig:initial_conditions}(b) at $\nu =10^{-6}$ and cutoff scale $\Delta=\ell/64$ for DNS (a), the NGMR model (b), DS model (c), and DM model (d). All fields were calculated from DNS data.}
    \label{fig:fluxes}
\end{figure*}

\section{Results}
To determine whether the SGS stress tensor can be represented by functions of the resolved fields $\bar{\bf u}$ and $\bar{p}$, we use them to construct the respective rank-0 and symmetric trace-free rank-2 term libraries and employ inhomogeneous regression. Inhomogeneous regression searches for sparse relations in the form $\tau=f(\bar{\bf u},\bar{p})$ with the known left-hand-side. Across all data sets, we consistently find the nonlinear gradient model (NGM2)
\begin{align}
    \tau_{ij} &\approx  c_1(\nabla_k\bar{u}_i)(\nabla_k\bar{u}_j)\equiv \tau_{ij}^{(2)}.
\end{align}
Using dimensional analysis and the numerical values of the coefficients identified by SPIDER for different data sets, we find $c_1=\Delta^2/12$, consistent with the moment expansion of $L$.  
To obtain a more accurate approximation, we repeat the inhomogeneous regression for the difference $\tau_{ij}-\tau_{ij}^{(2)}$, which 
consistently yields the next order correction
\begin{align}
    \tau_{ij}-\tau_{ij}^{(2)} &\approx 
    c_2 (\nabla_k\nabla_m\bar{u}_i)(\nabla_k\nabla_m\bar{u}_j)\equiv \tau_{ij}^{(4)}.
    \label{eq:ngm4}
\end{align}
Dimensional analysis and scaling yield $c_2=\Delta^4/288$, consistent with the moment expansion of $L+C-\tau^{(2)}$. Note that, for small $\Delta$,  $\tau^{(4)}$ becomes small compared with $\tau^{(2)}$ and, as a result, is generally omitted by inhomogeneous regression for $\tau$. The iterative procedure described here reliably picks up this term. We will refer to the parameterization $\tau=\tau^{(2)}+\tau^{(4)}$ as the NGM4 model.

As we show later, NGM4 is still not capable of accurately describing the energy flux. Hence, we repeat the iterative procedure to identify the next, even smaller correction $\tau^{(6)}=O(\Delta^6)$. However, this time we find no robust parameterization for $\tau^{(6)}$ in terms of $\bar{\bf u}$ and $\bar p$ with even moderate accuracy, which means that $\tau^{(6)}$ must be expressed in terms of an additional tensor field describing the small scales. Since NGM4 already accounts for the contributions from $L$ and $C$ in the decomposition \eqref{eq:LCR}, it is natural to assume that $\tau^{(6)}\approx R$ and treat $R$ as an additional variable in the EFT. This leads to the following parameterization of the SGS tensor
\begin{eqnarray}\label{eq:ngmr}
    \tau \approx \tau^{(2)} + \tau^{(4)} + R
\end{eqnarray}
which we refer to as the NGMR model.
This model not only reproduces the individual components of the SGS stress tensor faithfully but also accurately captures the spatiotemporal structure of the energy flux $\Pi$ \textit{including} backscatter (regions where $\Pi<0$) as illustrated in figure \ref{fig:fluxes} for a particular flow and a particular choice of viscosity and the filter scale. For reference, the figure also shows the predictions of the widely used dynamic Smagorinsky (DS) model as well as the dynamic mixed (DM) model (see supplemental material \cite{supplemental_material}), which is considered to be one of the most accurate LES models \cite{meneveau2000}.

In fact, the NGMR model succeeds on every metric and in every case we considered, while the phenomenological models rarely do. 
To measure the accuracy of an LES model in predicting a tensor variable, we employ a magnitude-aware correlation
\begin{align}
    \mathcal{C}(A,B)=\frac{\langle A_{ij\cdots}B_{ij\cdots}\rangle}{\max\left(\langle A_{ij\cdots}A_{ij\cdots}\rangle,\langle B_{ij\cdots}B_{ij\cdots}\rangle\right)}
\end{align}
between tensor fields $A$ (ground truth) and $B$ (prediction), where $\langle\cdot\rangle$ denotes spatial \textcolor{black}{and temporal averaging over the whole numerical solution}, with $\mathcal{C}(A,B)=1$ precisely when $A=B$.
In particular, $C_\tau= C(\tau_{\rm DNS},\tau_{\rm LES})$ and  $C_\Pi= C(\Pi_{\rm DNS},\Pi_{\rm LES})$ describe the accuracy in predicting the components of the SGS tensor and the energy flux, respectively. The accuracy in predicting the net energy flux can be described in terms of the ratio $q_\Pi= \langle\Pi_{\text{LES}}\rangle / \langle\Pi_{\text{DNS}}\rangle$. All three quantities should be equal to unity (100\%) if the prediction is perfect. As table \ref{tab:sample} shows, both phenomenological models fail to predict the local fluxes for any flows and the net flux for flows F1 and F3. An unexpected observation is that a very accurate prediction of $\tau$ (as in the case of NGM4) does  not guarantee accuracy in predicting the local and global energy fluxes. NGMR is the only model capable of predicting all these quantities across the board.

\begin{table*}[!htbp]
\centering
\begin{tabular}{c|c|ccc|ccc|ccc}
 &$\nu$ & & $10^{-4}$ & & & $10^{-5}$ & & & $10^{-6}$ & \\ 
 & $Re$ & $2\times 10^4$ & $2\times 10^4$  & $3\times 10^5$ & $3\times 10^5$ & $2 \times 10^5$ & $3 \times 10^6$ & $2 \times 10^6$ & $2\times 10^6$ & $3 \times 10^7$\\
\midrule

 & model & F1 & F2 & F3 & F1 & F2 & F3 & F1 & F2 & F3\\
\midrule
& DS & 0.13\% & 0.01\% & 0.01\% & 0.24\% & 0.01\% & 0.01\% & 0.27\% &  0.01\% & 0.01\% \\
& DM & 82.28\% & 99.71\% & 99.67\% & 73.52\% & 96.25\% & 99.67\% & 71.44\% & 96.17\%  & 99.66\%\\
$C_\tau$ & NGM2  & 92.27\% & 99.27\% & 99.94\% & 89.21\% & 99.15\% & 99.93\% & 88.44\% & 99.14\%  & 99.934\% \\
& NGM4  & 97.70\% & 99.93\% & 99.99\% & 95.94\% & 99.89\% & 99.99\% & 95.40\% & 99.89\%  & 99.99\%\\
& NGMR & 99.47\% & 99.98\% & 99.99\% & 99.34\% & 99.97\% & 99.99\% & 99.33\% & 99.97\%  & 99.99\%\\
\midrule
& DS & 26.01\% & 24.86\% & 22.77\% & 26.28\% & 27.87\% & 24.05\% & 22.48\% & 30.63\%  & 24.06\%\\
& DM & 27.62\% & 18.00\% & 15.64\% & 29.63\% & 20.20\% & 13.41\% & 27.03\% & 21.08\% & 13.40\% \\
$C_\Pi$ & NGM2  & 0\% & 0\% & 0\% & 0\% & 0\% & 0\% & 0\% &  0\% & 0\% \\
&  NGM4   & 83.49\% & 90.57\% & 99.44\% & 75.41\% & 83.81\% & 94.12\% & 72.26\% & 83.07\%  & 93.98\% \\
& NGMR  & 96.85\% & 98.02\% & 99.85\% & 96.97\% & 97.52\% & 98.89\% & 97.12\% & 97.40\% & 98.83\% \\
\midrule
& DS & 131.76\% & 107.25\% & 31.53\% & 129.05\% & 106.27\% & 26.87\% & 128.74\% & 106.17\%  & 26.86\% \\
& DM & 134.12\% & 109.92\% & 35.03\% & 131.36\% & 108.93\% & 30.19\% & 131.04\% & 108.83\%  & 30.18\% \\
$q_\Pi$ & NGM2  & 0\% & 0\% & 0\% & 0\% & 0\% & 0\% & 0\% & 0\% & 0\% \\
 & NGM4  & 59.31\% & 69.83\% & 82.33\% & 49.13\% & 62.29\% & 58.60\% & 46.33\% & 60.78\% & 56.28\% \\
& NGMR & 104.18\% & 104.63\% & 103.38\% & 103.81\% & 104.08\% & 104.94\% & 102.71\% & 103.93\% & 104.63\% \\

\end{tabular}
\caption{The accuracy, averaged in time over the dataset, of the SGS stresses and corresponding energy fluxes for flows F1--F3, with $R$ computed from DNS. To test the limits of the inferred parameterization, we purposely chose a coarse resolution with a cutoff scale $\Delta=\ell/64$ that is close to the size of the eddies in the flow F1. 
}
\label{tab:sample}
\end{table*}

\begin{figure}[htbp!]
    \centering
    \includegraphics[width=0.4\textwidth]{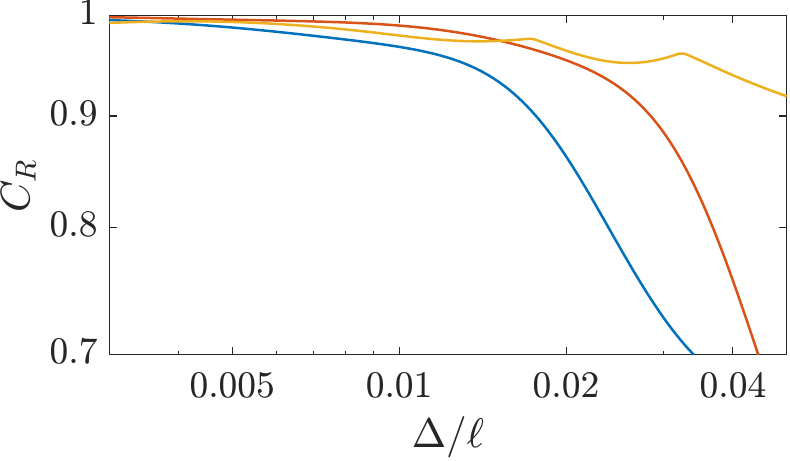}
    \caption{
        The accuracy of the evolution equation \eqref{eq:R}, quantified by the correlation $C_R$, as a function of  $\Delta/\ell$
        for F1 (blue), F2 (orange) and F3 (yellow)
        at $\nu = 10^{-6}$.}
    \label{fig:R_res}
\end{figure}

The inclusion of $R$ as a new field requires an additional governing equation in order to close the system. This equation can be inferred using rank-0 and symmetric trace-free rank-2 libraries constructed from $\bar{u}_i$, $\bar{p}$, and $R_{ij}$ and homogeneous regression, which searches for sparse relations in the form $f(R,\bar{\bf u},\bar{p})=0$. We robustly discover an evolution equation of the form
\begin{align}
\partial_tR_{ij} & \approx c_1\bar{u}_k\nabla_k R_{ij} +c_2(R_{ik}\nabla_k \bar{u}_j + R_{jk}\nabla_k \bar{u}_i)  \nonumber\\
& + c_3\nabla^2R_{ij}+c_4 R_{ij}.\label{eq: R evo}
\end{align}
Using a combination of dimensional analysis and numerical values of the coefficients for different data sets, we find $c_1=-1$, $c_2=1$ and $c_3=\nu$.
For $c_4$, we find that it has both dimensions and numerical values consistent with the inverse of the eddy turnover time but exhibits no clear dependence on the parameters $\nu$ and $\Delta$.
Therefore, we relax SPIDER's assumption of translational invariance that requires $c_4$ to be independent of position and time and take it to be a function of the resolved flow. For instance, in the inviscid limit, rescaling the velocity of the initial condition by a constant factor changes the inverse time scale and, hence, $c_4$ by the same factor, so $c_4$ should be linear in $\bar{u}$. Because $c_4$ is a scalar, it should be expressible as a function of the energy $E=u_iu_i/2$ as well as the invariants of the tensor $\nabla_i\bar{u}_j$ \cite{pope1975}; in 2D, these are given by $|\bar{S}|=\sqrt{2 \bar{S}_{ij}\bar{S}_{ij} }$ and $|\bar{\Omega}|=\sqrt{2\bar{\Omega}_{ij} \bar{\Omega}_{ij}}$, where ${\Omega}_{ij}$ is the antisymmetric component of $\nabla_i u_j$.
Dimensional arguments would then require $c_4=\Delta^{-1}\sqrt{E}f(\Delta|\bar\Omega|/\sqrt{E},\Delta|\bar S |/\sqrt{E},\cdots)$, where $f$ is an arbitrary function.
A general relation of this form can be inferred from the data using, e.g., symbolic regression. However, a good approximation can be found using a series expansion of $f$: 
\begin{eqnarray}
    c_4 =  \alpha_1 |\bar S| + \alpha_2 | \bar \Omega | + \alpha_3\Delta^{-1}\sqrt{E} + \cdots,
\end{eqnarray}
where $\alpha_j$ are non-dimensional coefficients. Sparse regression yields $c_4 \approx -|\bar S|/2$. 
Collecting everything, we find
\begin{eqnarray}\label{eq:R}
\partial_tR_{ij} & +\bar{u}_k\nabla_k R_{ij} \approx R_{ik}\nabla_k \bar{u}_j + R_{jk}\nabla_k \bar{u}_i  \nonumber \\
& + \nu\nabla^2R_{ij}-\frac{1}{2} |\bar S|R_{ij}
\end{eqnarray}
Using $C_R=\mathcal{C}(\partial_tR_{\text{DNS}},\partial_tR_{\text{NGMR}})$ as a metric, we find equation \eqref{eq:R} to accurately describe all flow regimes for sufficiently small $\Delta$ (cf. figure \ref{fig:R_res}).
As expected, the accuracy deteriorates as $\Delta$ approaches the scale of coherent structures, since the coarse-grained description is not meant to model these.

\subsection*{DISCUSSION}
{\color{black}Taken together, the closed system of equations \eqref{eq:les}, \eqref{eq:ngmr} and \eqref{eq:R} defines an EFT which describes incompressible fluid turbulence up to the cutoff scale $\Delta$. 
\textcolor{black}{It should be emphasized that, since the same equations with the same coefficients are inferred from nine different data sets, the form of the EFT is quite general and independent of our choice of the flows F1, F2, and F3. On the other hand, since the training data represents $Re> 10^4$, it may or may not extend to lower $Re$.}
By construction, this theory is explicitly equivariant with respect to translations and rotations but also turns out to be explicitly Galilean-invariant, as it should. Not only is the resulting description more accurate and general than state-of-the-art LES models---at least in two spatial dimensions---it is explicit and easily interpretable, which allows one to make specific predictions. For example, at scales $\Delta$ small compared to the size of large-scale coherent structures, the energy flux is dominated by the contribution from the $C$ tensor and scales as $\Delta^4$, which is not directly implied by the fundamental model \eqref{eq:fm}. Our data-driven framework also identifies two methodological problems with traditional modeling approaches. Accurate description of inter-scale fluxes requires introduction of additional fields describing subgrid scales which are missing in all LES models. Moreover, these fields must have a tensor structure, unlike the most common RANS models such as $k-\epsilon$, $k-\omega$, and Spalart-Allmaras, which all employ scalar fields \cite{launder1974, wilcox2008, spalart1992}.}

\textcolor{black}{The results presented here only validate the NGMR model in an {\em a priori} sense, with every model variable evaluated using a filtered DNS solution. An {\em a posteriori} validation---which does not rely on the DNS solution--- requires a numerical implementation of all of the coupled equations comprising an EFT. Therefore, while the presented model yields useful theoretical insights, it can falter in numerical implementations. Indeed, a numerical implementation's accuracy or even stability is not guaranteed, whether for phenomenological or data-driven SGS models, and often requires additional modifications. For instance, the dynamic Smagorinsky model, both in standalone form or as a part of the dynamic mixed model, typically employs clipping of the backscatter regions, i.e., setting $\tau_{ij}=0$ wherever $\Pi<0$, just to ensure stability. While the focus of this paper is on methodology rather than {\em a posteriori} validation of the NGMR model, we emphasize that the {\em a priori} accuracy of a SGS model is a prerequisite for its {\em a posteriori} accuracy \cite{moser2021}. 
}

\textcolor{black}{Given that the SGS modeling framework did not rely on any phenomenological assumptions, it is natural to ask whether it would generalize to multiscale phenomena other than turbulence in incompressible fluids.}
\color{black}
Consider a system described by a finite number of tensor fields (tensorial indices suppressed) $q^{(\alpha)}$, ${\alpha} = 1,2,\cdots$ with evolution equations
\begin{eqnarray}\label{eq:genev}
    \partial_t q^{{(\alpha)}} + \sum_k N_k^{(\alpha)}(q) = S^{(\alpha)}(q),
\end{eqnarray}
where $q \equiv \{q^{(1)},q^{(2)},\cdots\}$, $S^{(\alpha)}$ are terms linear in $q$, and $N_k^{(\alpha)}$ are quadratic nonlinearities. We will assume that $N_k^{(\alpha)} =D_k^{(\alpha)}(A^{(\alpha)}_kB_k^{(\alpha)})$, where $D_k^{(\alpha)}$ are some differential operators acting on tensor products of linear functions $A_k^{(\alpha)}(q^{(\gamma)})$ and $B_k^{(\alpha)}(q^{(\beta)})$ with arbitrary $\gamma$ and $\beta$.  Examples of such nonlinearities are the terms $\nabla\cdot({\bf u}\otimes {\bf u})$ in the momentum equation or $\nabla\times({\bf u}\times {\bf B})$ in the induction equation in magnetohydrodynamics. Filtering equations \eqref{eq:genev} yields
\begin{eqnarray*}
    \partial_t \bar q^{(\alpha)} +  \sum_k  N_k^{(\alpha)}(\bar q) = S^{(\alpha)}(\bar q) - \sum_k D_k^{(\alpha)} Q_k^{(\alpha)}(q,\bar q),
\end{eqnarray*}
where $D_k^{(\alpha)} Q_k^{(\alpha)}$ are the closure terms with closure variables
\begin{eqnarray*}
    Q_k^{(\alpha)} = \overline{A_k^{(\alpha)} B_k^{(\alpha)}}-\bar A_k^{(\alpha)} \bar B_k^{(\alpha)}.
\end{eqnarray*}
Each tensor $Q^{(\alpha)}_k$ can be decomposed, similar to Equation \eqref{eq:decomp}, into 
components 
\begin{eqnarray*}              Q_k^{(\alpha)} &=& L_k^{(\alpha)}+C_k^{(\alpha)}+R_k^{(\alpha)} 
    \\
    L_k^{(\alpha)} &=&   \overline{\overline A_k^{(\alpha)} \overline B_k^{(\alpha)}} - \overline{ \overline{A}}_k^{(\alpha)} \overline{\overline B}_k^{(\alpha)}
   \\
      C_k^{(\alpha)}  &=&  \overline{\overline{A}_k^{(\alpha)} {B'}^{(\alpha)}_k + \overline{B}_k^{(\alpha)} {A'}^{(\alpha)}_k } -\overline{\overline A}_k^{(\alpha)} \overline{B'}_k^{(\alpha)} - \overline{\overline B}_k^{(\alpha)} \overline{A'}_k^{(\alpha)} 
   \\
  R_k^{(\alpha)}  &=&  \overline{ {A'}_k^{(\alpha)} {B'}_k^{(\alpha)} } - \overline{A'}_k^{(\alpha)} \overline{B'}_k^{(\alpha)}
\end{eqnarray*}
describing, respectively, the interactions between resolved scales, resolved and subgrid scales, and subgrid scales. The gauge symmetry
\begin{eqnarray*}
    Q_k^{(\alpha)}(A_k^{(\alpha)}+a,B_k^{(\alpha)}+b) = Q_k^{(\alpha)}(A_k^{(\alpha)}, B_k^{(\alpha)})
\end{eqnarray*}
where $a$ and $b$ are any spatially uniform tensor fields, is inherited individually by each of the three components.
In the case of the momentum equation, this symmetry corresponds to Galilean invariance. 

Generally, each of the $L_k^{(\alpha)},C_k^{(\alpha)},R_k^{(\alpha)}$ tensors must be approximated well for $Q_k^{(\alpha)}$ to be accurately modeled. To this end, we distinguish tensors that are \textit{resolvable}---those that can be parameterized effectively in terms of the respective $\bar{q}^{(\gamma)}$ and $\bar{q}^{(\beta)}$ with the moment expansion on the coarse grid---and \textit{unresolvable}---those that cannot.
This notion of resolvability generalizes to arbitrary systems (with quadratic nonlinearities), as the moment expansion is a formal series expansion and does not rely on the system. Moreover the hierarchical scaling $L_k^{(\alpha)} = O((\Delta/\ell_c)^2),\ C_k^{(\alpha)} = O((\Delta/\ell_c)^4),\ R_k^{(\alpha)} = O((\Delta/\ell_c)^6)$ holds generally, too. $L^{(\alpha)}_k$ is a function of resolved scales by definition and hence requires no explicit modeling---one can use either its moment expansion or its definition. This is the path conventional SGS models tend to follow while ignoring contributions from $C_k^{(\alpha)}$ and $R_k^{(\alpha)}$.

$C_k^{(\alpha)}$ and $R_k^{(\alpha)}$ generally require modeling, since both components formally depend on the subgrid scales. However, it may be the case that either (or both) can be parameterized effectively with the moment expansion. The resolvability of each $C_k^{(\alpha)}$ and $R_k^{(\alpha)}$ can be determined by performing iterative inhomogeneous regression and inspecting the structure of the inferred parameterizations and the associated residuals. It is possible for all $L^{(\alpha)}_k,C^{(\alpha)}_k, R^{(\alpha)}_k$
tensors to be parameterized by the moment expansion with adequate accuracy if the spectra of the respective tensors decay quickly enough. In such a case, no additional equations are needed to close the system. Note that, even if $R_k^{(\alpha)}$ is found to be unresolvable in terms of $\bar{q}^{(\gamma)}$ and $\bar{q}^{(\beta)}$, it may still be possible to express $C_k^{(\alpha)}$ in terms of $\bar{q}^{(\gamma)}$, $\bar{q}^{(\beta)}$ and $R_k^{(\alpha)}$.  
\textcolor{black}{It should be cautioned that it is not guaranteed that accurate evolution equations (or parameterizations) for all the modeled variables can be found. This is the largest uncertainty inherent in this framework.}

We should emphasize a practical issue that is critical to the success of data-driven modeling but can be easily overlooked.
It is of utmost importance that all the variables are projected onto the coarse grid before any functional relations are inferred using SPIDER; it is insufficient to only filter the variables. Indeed, the filtering operation is invertible in the spatially continuous case (or on a grid sufficiently fine to resolve the solutions of the fundamental equations). Specifically, the moment expansion can be used to recover the original variables---and hence the subgrid scales---from their filtered versions with high accuracy on a fine grid. As a consequence of this (as well as the fast scaling of the $C$ and especially $R$ component of closure variables with $\Delta$), testing SGS models on fine grids is rather meaningless and should be conducted on grids substantially coarser than those needed to fully resolve the solutions of the fundamental equations. 

Once the unresolvable tensors are ascertained, these tensors should be treated as new variables in the EFT. The corresponding evolution or constitutive equations necessary to close the system can be inferred using homogeneous regression. These equations are inferred with specific numerical values for each of the coefficients (i.e., parameters of the EFT). A combination of dimensional analysis and scaling can be used to identify the functional relation between the EFT parameters and the parameters of the fundamental equations \eqref{eq:genev}. 
As the example of fluid turbulence illustrates, the coefficients of the inferred evolution/constitutive equations can also depend on various tensor invariants. For example, some parameters of the evolution/constitutive equation for $R^{(\alpha)}_k$ can depend on the invariants $I^{(\alpha)}_{k,\eta}$ of the tensors $\bar{q}^{(\eta)}$, $\Delta\nabla_i\bar{q}^{(\eta)}$, $\Delta^2\nabla_i\nabla_j\bar{q}^{(\eta)}$, etc. with $\eta=\gamma,\beta$ corresponding to the particular $k$ and $\alpha$.
The scaling of the invariants with $\Delta$ shows that only a finite number of these (corresponding to the lowest powers of $\Delta$) need to be considered. The functional dependence on these invariants can be inferred using either sparse regression or, failing that, symbolic regression.

\color{black}
\section{Conclusions}

{\color{black}This paper introduces a general SGS modeling framework that combines elements of formal analysis and machine learning. The framework is illustrated by using filtered DNS data to infer a symbolic SGS model of 2D incompressible fluid turbulence that outperforms state-of-art LES models on a variety of {\em a priori} benchmarks. }
The traditional modeling approach has required decades of slow iterative progress and numerous---and often poorly justified---phenomenological assumptions to develop the current generation of LES models. In contrast, the framework presented here is both far more efficient and produces more accurate SGS models. Most importantly, it requires a minimal number of ingredients: the fundamental model (which could also be derived in a data-driven way \cite{gurevich2024}), some of the fundamental symmetries of the problem, dimensional analysis, and no specific phenomenological assumptions. Hence, it {\color{black}is expected to} generalize to other multiscale problems described by equations with quadratic nonlinearities.

\textit{Acknowledgments---}This work was supported by the Defense Advanced Research Projects Agency (DARPA). {\color{black}This research was also supported in part through research cyberinfrastructure resources and services provided by the Partnership for an Advanced Computing Environment (PACE) at the Georgia Institute of Technology, Atlanta, Georgia, USA.}

\end{document}